\documentclass[secnumarabic, graphics,floatfix,nofootinbib,
tightenlines,nobibnotes,aps,prl,10pt]{revtex4-1}
\usepackage{amsmath}
\usepackage{graphicx}
\usepackage{multirow}
\usepackage{listings}

\begin{document}

\begin{titlepage}
    \begin{center}
        {\LARGE A Nonconventional Analysis of CD$4^{+}$ and CD$8^{+}$ T Cell Responses During and After Acute Lymphocytic Choriomeningitis Virus Infection \par}
        \vskip 2em
        { \small by} \\
        Dwayne John \\
        \vskip 2em
        \vskip 1em
        {\Large \makebox[3in]{\hrulefill} \par}
        \vskip 1em
        {\small
            \today \\
            Computational Science Program\\
            Department of Chemistry\\
            Middle Tennessee State University\\
            301 E Main St\\
            Murfreesboro, TN 37132\\}
    \end{center}
    \par
    \global\let\newpagegood\newpage
    \global\let\newpage\relax
\end{titlepage}
\global\let\newpage\newpagegood

\setcounter{secnumdepth}{5} 
 \setcounter{tocdepth}{5} 

\noindent A mathematical model from a previous work was re-fitted and analyzed for experimental data regarding the cellular immune response to the lymphocytic choriomeningitis virus. Specifically, the $CD8^{+}$ T cell response to six MHC class I-restricted epitopes (GP* and NP*) and $CD4^{+}$ T cell responses to two MHC class II-restricted epitopes\cite{de2003different}. In this work, we use calibration through log likelihood maximization to investigate if different parameters can produce a more accurate fit of the model presented previously in the paper titled \textit{Different Dynamics of CD$4^{+}$  and CD$8^{+}$  T Cell Responses During and After Acute Lymphocytic Choriomeningitis Virus Infection}\cite{de2003different}. 

\newpage

 \section{Historical Background}
\noindent The leading cause of death among people under the age of 45 is trauma. A main cause of death after trauma is internal bleeding of the abdominal organs. Of those abdominal organs, the spleen is one of the most often injured due to blunt trauma\cite{vanderVlies2011}, so much so that it is affected in 32\% of patients that have traumatic abdominal injuries\cite{van2012impact}. Every year, about 39,000 patients are admitted in hospitals throughout the United States of America for the treatment of blunt splenic trauma (BSI). Of those admitted patients, 28,000 will undergo nonoperative management care\cite{requarth2010distal}. \\
\\
It was once believed that the spleen could not heal spontaneously and would also rupture later during patient recovery. These ideas started to be re-examined in the 1970s when data about post-operative infections regarding laparotomies and removing the spleen was published\cite{vanderVlies2011}. Due to this new knowledge, more doctors are now performing procedures such as splenic artery embolization (SAE) which preserve the spleen and also hold a high success rate\cite{van2012impact}.\\
\\
Like the appendix, the spleen's function in the human body was unknown to many. In 1919, two individuals (Morris and Bullock) showed that dogs with their spleen that were infected with rat plague bacillus had a higher survival rate as opposed to those that were infected without their spleen. Due to limitations in diagnostic medicine at the time, splenectomy was a main standard of care. Until about 65 years ago, no one even questioned the idea of splenectomy as a treatment en masse when there were a plethora of overwhelming polstsplenectomy infection (OPSI) cases. Things started to change roughly fifty years ago when care shifted to splenorraphies and then again about a decade later when clinicians started expectant management of patients became mainstream. Contrary to older ideas regarding the spleen's function, we now know the spleen assists with immune function. It can help filter specific antigens and microorganisms and act in cell regeneration\cite{kaseje2008short}. 

\section{Introduction}

\noindent The lymphocytic choriomeningitis (LCM) virus is a rodent disease that an be passed in many different ways from rodent to rodent, human to human, and rodent to human. It can stay dormant in mice for as long as 35 years\cite{lehmann1971lymphocytic}. A member of the arenaviridae family of viruses, lymphocytic choriomeningitis virus (LCMV) is a human pathogen that can infect a large amount of the human population\cite{bonthius2012lymphocytic}.


\section{Materials and Methods}

\noindent In the previous study by De Boer et al.\cite{de2003different}, data was collected by first injecting six to eight week old mice, male and female, with LCMV. At different time intervals, spleen cell samples were extracted and measured from an average of three to four mice per data point.  The number of specific T cells per spleen was measured. A set of linear differential equations was used to describe the data. Due to the linear nature of the model, a solution to the differential equations can be obtained. De Boer, et al. estimated the parameters of the model using the DNLS1 subroutine from the Common Los Alamos Software Library and based on the Levenberg-Marquardt algorithm\cite{more1978levenberg}.\\
\\
We employ a different, more efficient strategy to arrive at the same destination: parameter optimization through log likelihood maximization.\\

\begin{figure}[h]
\caption{Cubic spline interpolation of specific $CD4^{+}$ T cells per spleen (two responses GP61 and NP309) with respect to days after LCMV infection\cite{de2003different}.}
\centering
\includegraphics[width=10cm]{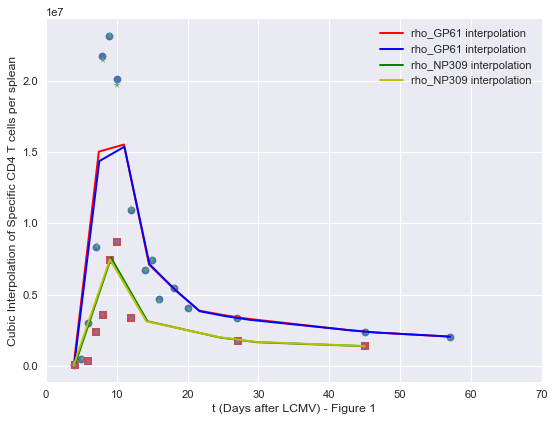}
\end{figure}

\begin{figure}[h]
\caption{Cubic spline interpolation of specific $CD4^{+}$ T cells per spleen (two responses GP61 and NP309)  with respect to days after LCMV infection\cite{de2003different}.}
\centering
\includegraphics[width=10cm]{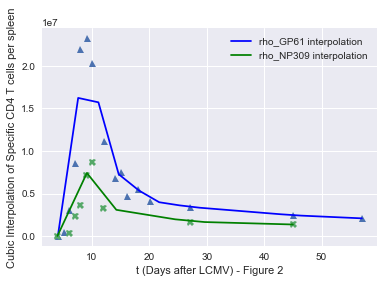}
\end{figure}

\begin{figure}[h]
\caption{Cubic spline interpolation of specific $CD8^{+}$ T cells per spleen (six responses GP33, GP118, NP205, NP396, GP276, GP92) with respect to days after LCMV infection\cite{de2003different}.}
\centering
\includegraphics[width=10cm]{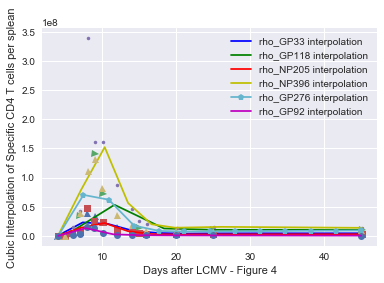}
\end{figure}

\noindent In order to calculate the model results with the necessary parameters, we first calculate the values of the population of activated cells, A and the memory cells, M. The death rate of memory cells and activated cells is given by $\delta_M$ and $\delta_A$, respectively. \\
\\
There are three phases in the model: rapid expansion, rapid contraction, and slower contraction of the T cell population.
\noindent The three phases for modeling the amount of T cells per spleen correspond to the following time intervals: t $<$ T,  t between T and T+$\Delta$, and T$>$T+$\Delta$.\\
\\
The population of A cells in the initial phase of rapid cellular growth is given by the following differential equation:
$$\frac{dA}{dt} = \rho A,$$
the parameter $\rho$ is the net expansion rate.\\ 
\\
The population of A and M cells during the contraction phase is given by the following equations:
$$\frac{dA}{dt} = -(r + \alpha + \delta_A)A$$\\
\\
\\
$$\frac{dM}{dt} = rA - \delta_M M$$\\
\\
\\
The parameter $\alpha$ is the cell death rate, also known as rapid apoptosis and $r$ is the contraction phase rate. At time T, A reaches its peak value and is calculated by $A(T) = A(0)\exp[\rho T]$.\\
\\
We then calculate the value of A at T + $\Delta$. The $\Delta$ term is the duration of the phase of rapid contraction of T cells population. That phase is followed by a phase of slower contraction of the population. Now, $A(T+\Delta) = A(T)\exp[-\delta(\Delta)]$. Next, we calculate the value of M at T+$\Delta$, $$M(T+\Delta) =  \frac{\exp[-\delta_M(T+\Delta)](r A(T)(1-\exp[ -(\delta_{A}-\delta_{M}) (T+\Delta) ] ) )}{\delta-\delta_M}$$


\noindent For $t < T$, $A = A(0)\exp[\rho t]$ and M=0.\\
For $t > T$, $A = A(T)\exp[-\delta(t-T)]$ and $$M = \frac{\exp[-\delta_M(t)]r A(T) (1-\exp[-(\delta-\delta_M)t])}{\delta-\delta_M},$$ \\

where $\delta=r+\delta_{A}+\alpha$.\\

\noindent For $T < t < T + \Delta$, $A = A(T + \Delta)\exp[-\delta^\prime(t-T-\Delta)]$ and $$M(t) = \frac{\exp[-\delta_M t](rA(T+\Delta)(1-\exp[-(\delta^\prime-\delta_M)t]))+M(T+\Delta)(\delta^\prime-\delta_M)}{\delta^\prime-\delta_M},$$

and $\delta^{\prime} = r + \delta_{A}$.\\

\noindent Data was collected from citation  \#\cite{de2003different} using the PlotDigitizer application.  A Macbook Pro running Anaconda Python 3.6 with the Jupyter Notebooks was used for all calculations.  

\subsection{Sensitivity Analysis}

\noindent Sobol method was used first to see which parameters have the most influence on the fluctuations of the results. The results of these calculations were inconclusive and we were not able to ascertain which parameter had the greatest effect. We suspect that all parameters have a similar influence on the results of the differential equations.




\section{Discussion}

\noindent Our sensitivity analysis on the parameters of the model did not show that any particular parameter is significantly more important than the others. \\
\\
It was found that the optimized parameters calculated in this work, using a simple log likelihood maximization method, are noticeably different from those calculated in the paper\cite{de2003different}, using the Levenberg-Marquard algorithm. \\
\\
A few factors that influenced the results are the following. The data was acquired directly from the graphs in the published paper\cite{de2003different} and not provided via a database from the corresponding author due to time constraints of the project. The use of this method is subject to errors that are difficult to quantify due to human-machine interaction. Another influential factor is that our study used only the data corresponding to the first 70 days, while the original article used the full data set up to 921 days. Despite these issues, in most cases the newly estimated parameters fall within the 95\% combined confidence intervals reported in the paper.\\


\begin{table}[h]
\centering
\begin{tabular}{|l|c|c|c|c|c|}
\hline
 \bf Parameter & \bf Log Likelihood Maximization Method & \bf 95\% Combined CI\cite{de2003different} & \bf Units \\
\hline
$\rho$ & 1.82 &0.98-2.21 &  $d^{-1}$\\
\hline
$\delta_{A}$ & 0.01182 & 0.01-0.47 & $d^{-1}$ \\
\hline
$\delta_{M}$ & 0.00017 & 0.0006-0.003 &  $d^{-1}$ \\
\hline
$r$ & 0.00094 & 0.001-0.022 & $d^{-1}$ \\
\hline
$\alpha$ & 0.16750 &  0.12-0.82 & $d^{-1}$\\
\hline
$T$ & 7.72 & 8-9.1 & Days \\
\hline
$\Delta$ & 10.4 & 2.2-9.8 & Days \\
\hline
$A(0)$ & 19.3 & 0.4-518.8 & Cells \\
\hline
\end{tabular}
\caption{Results for implementation of log-likelihood maximization, compared to De Boer et al.'s results from the paper\cite{de2003different}.}
\label{tab:template}
\end{table}

\section{Conclusion}

\noindent We can conclude that, in cases where a quick and simple verification is in order, ``data thieving'' from published graphs and the log likelihood maximization method can be useful tools. In this case, we were able to obtain results consistent with previously published\cite{de2003different} work using the calibration likelihood estimation parameter optimization. 

\listoffigures
\listoftables

\newpage

\newpage
\bibliographystyle{aipauth4-1}
\bibliography{bibl}

\end{document}